\documentclass[preprint2]{aastex}

\shorttitle{A New Collisional Ring Galaxy in Auriga}
\shortauthors{Blair Conn et al.}

\begin{document}


\title{A New Collisional Ring Galaxy at {\it z}  = 0.111: Auriga's Wheel.}


\author{Blair C. Conn\altaffilmark{1},\email{conn@mpia-hd.mpg.de} Anna Pasquali\altaffilmark{2},\email{pasquali@ari.uni-heidelberg.de} Emanuela Pompei\altaffilmark{3},\email{pompei@eso.org} Richard R. Lane\altaffilmark{4}, \email{rlane@astro-udec.cl} Andr\'{e}-Nicolas Chen\'{e}\altaffilmark{4,5},\email{achene@astro-udec.cl} Rory Smith\altaffilmark{4} \email{rsmith@astro-udec.cl} and Geraint F. Lewis\altaffilmark{6} \email{geraint.lewis@sydney.edu.au}}
\altaffiltext{1}{Max Planck Institut f\"{u}r Astronomie, K\"{o}nigstuhl 17, 69117, Heidelberg, Germany}
\altaffiltext{2}{Astronomisches Rechen-Institut, Zentrum f\"{u}r Astronomie der Universit\"{a}t Heidelberg, M\"{o}nchhofstrasse 12-14, 69120,Heidelberg, Germany}
\altaffiltext{3}{European Southern Observatory, Alonso de C\'{o}rdova 3107, Casilla 19001, Vitacura, Santiago, 19, Chile}
\altaffiltext{4}{Departamento de Astronom\'{i}a, Universidad de Concepc\'ion, Casilla 160-C, Concepc\'ion, Chile}
\altaffiltext{5}{Departamento de F\'isica y Astronom\'ia, Facultad de Ciencias, Universidad de Valpar\'{a}iso, Av. Gran Breta\~{n}a 1111,Playa Ancha, Casilla 5030, Valpara\,iso, Chile}
\altaffiltext{6}{Sydney Institute for Astronomy, University of Sydney, A28, Sydney, 2006, Australia}


\begin{abstract}
We report the serendipitous discovery of a collision ring galaxy, identified as  2MASX J06470249+4554022, which we have dubbed `{\it Auriga's Wheel}', found in a SUPRIME-CAM frame as part of a larger Milky Way survey. This peculiar class of galaxies is the result of a near head-on collision between typically, a late type and an early type galaxy.  Subsequent GMOS-N long-slit spectroscopy has confirmed both the relative proximity of the components of this interacting pair and shown it to have a redshift of 0.111.  Analysis of the spectroscopy reveals that the late type galaxy is a LINER class Active Galactic Nuclei while the early type galaxy is also potentially an AGN candidate, this is very uncommon amongst known collision ring galaxies.  Preliminary modeling of the ring finds an expansion velocity of $\sim$200 km s$^{-1}$ consistent with our observations, making the collision about 50 Myr old. The ring currently has a radius of about 10 kpc and a bridge of stars and gas is also visible connecting the two galaxies. 
\end{abstract}





\section{Introduction}
Collisional ring galaxies (CRGs) are the result of a rare type of galaxy-galaxy interaction whereby one galaxy collides with the other along the axis of rotation. The induced gravitational instability generates a propagating density wave which sweeps the interstellar medium into an expanding star-forming ring ~\citep{Higdon2010}, the archetype for this galaxy class is the Cartwheel galaxy.  Currently, there are only about 127 spectroscopically confirmed CRGs in the literature \citep{Madore2009}, the majority of which are at relatively low redshift ($z = 0.002 - 0.088$). In the high redshift universe ($z = 0.4-1.4$) there are another 20 or so morphologically selected candidates having either photometrically or spectroscopically derived redshifts \citep{Lavery2004,Elmegreen2006}. All in all, this is a rare-type of encounter with a short lifespan of several hundred Myrs \citep{Moiseev2009,Higdon2010}, and so such small number statistics is not unexpected. Despite the scarcity of these galaxies, they are ideal laboratories for studying young stellar populations and stellar/gas dynamics within galactic disks.

As the ring of star formation ripples out from the point of impact the interstellar medium that it encounters is carried along by it forming new stars in the ring. As the wave of gas and dust continues on, these young stars are left behind and over the lifetime of the ring, new stars are continually created and left behind until no more stars can be formed or another event disrupts this star formation process. The final result is a radial gradient in the age of the stars in the galactic disk ~\citep{Bizyaev2007}, with old stars from the initial burst of star-formation residing at the centre and the youngest stars being formed in the ring. Comparing the velocity of the expanding ring with the ages of the young stellar populations present within the disk can be used to establish whether star formation was coincident or delayed with respect to the ring propagation. This provides insight into how star formation proceeds in more complicated systems like disk galaxies with spiral density waves. With regard to understanding gas and stellar dynamics, CRGs present an interesting yet tractable problem relating the initial gravitational instability and geometry of the interaction with the structures created. Broad overviews on CRGs can be found in ~\citet{Buta1996,Moiseev2009,Struck2010,Higdon2010} among others.

This work presents a new collisional ring galaxy (Figure~\ref{stampfig}) that has been found by visual inspection in two SUPRIME-CAM $g'$- and $r'$-band images obtained during a program to study the Milky Way Thick Disk. The prominent ring in the late-type galaxy and the apparent proximity to the early-type galaxy strongly suggested an ongoing interaction. Determining the $(g'-r')$ color and estimating a photometric redshift of z$\sim$0.18 ~\citep{Fukugita1995} brought the possibility that this galaxy pair would represent a high redshift collisional ring galaxy that can be studied in detail. With this in mind, we obtained time on the Gemini North telescope to perform long-slit spectroscopy and confirm the redshift while investigating the properties of the interaction. This pair of galaxies is located at RA: 06$\fh$47$\fm$2$\fs$3 and DEC: +45$\arcdeg$54$\arcmin$00$\arcsec$, (J2000.0) in the constellation of Auriga, the Charioteer (Table~\ref{xidtable}). With this in mind, coupled with it's wheel-like appearance, we have named it {\it Auriga's Wheel}.

\section{Observations and Data Reduction}
\subsection{Photometry}

\begin{figure}
\epsscale{.80}
\plotone{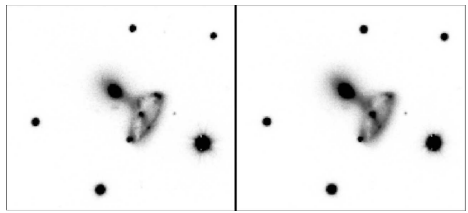}
\caption{SUPRIME-CAM stamp images of {\it Auriga's Wheel} in $g'$ (left) and $r'$ (right). The seeing was slightly better during the $g'$ band exposure making the structures within the interacting pair more visible. The $g'$ exposure is 124 seconds and the $r'$ exposure is 76 seconds. North is up, East is to the left. The images are $36.2 \times 32.6$ arcseconds.}\label{stampfig}
\end{figure}

\begin{deluxetable}{lccc}
\tabletypesize{\scriptsize}
\tablecaption{Coordinates of {\it Auriga's Wheel}.  \label{xidtable}}
\tablewidth{0pt}
\tablehead{
\colhead{Name} & \colhead{Cross-Identification} & \colhead{RA (2000)} & \colhead{DEC (2000)} 
}
\startdata
    ¥Auriga's Wheel& \nodata &06$\fh$ 47$\fm$ 02$\fs$3 & +45$\arcdeg$ 54$\arcmin$ 00$\arcsec$\\ 
    ¥Auriga's Wheel: N1& \nodata &06$\fh$ 47$\fm$ 02$\fs$1& +45$\arcdeg$ 53$\arcmin$ 59$\arcsec$\\ 
    ¥Auriga's Wheel: N2& 2MASX J06470249+4554022&06$\fh$ 47$\fm$ 02$\fs$5&+45$\arcdeg$ 54$\arcmin$ 02$\arcsec$\\ 
 \enddata
\end{deluxetable}

\begin{deluxetable}{cccc}
\tabletypesize{\scriptsize}
\tablecaption{Observations of {\it Auriga's Wheel}.  \label{obstable}}
\tablewidth{0pt}
\tablehead{
\colhead{Instrument} & \colhead{Filter/Grating} & \colhead{Exposure time} & \colhead{Date} 
}
\startdata
    ¥SUPRIME-CAM & ¥$\it{g'}$ & 124s¥ & ¥ 2007 Nov 9 \\ 
    ¥SUPRIME-CAM & ¥$\it{r'}$ & ¥76s & ¥  2007 Nov 9\\ 
    ¥GMOS-N & ¥R400 & ¥3$\times$1200s & ¥ 2009 Oct 31\\ 
    ¥GMOS-N& ¥R400 & ¥3$\times$1200s & ¥ 2009 Nov 6 \\ 
 \enddata
\end{deluxetable}

The data for the initial discovery of this collision ring galaxy was acquired with SUPRIME-CAM on the Subaru Telescope on 2007 November 9. Details of the data reduction process for the imaging dataset can be found in ~\citet{Conn2011b}. In brief, the photometry for {\it Auriga's Wheel}, comes from a single $g'$ and $r'$ band image from SUPRIME-CAM with exposure times of 124 and 76 seconds respectively (Figure~\ref{stampfig}). The pixel scale is 0.2 arcseconds per pixel and the seeing was 0.36 and 0.48 arcseconds in the $g'$ and $r'$ band respectively as measured from the image.  The magnitude of the stars in the frame were determined by using a Sloan Digital Sky Survey field for calibration of the color term and zeropoint. The conversion from flux to magnitudes used derived zeropoints of 27.67 mag in the $r$-band and 27.45 mag in the $g$-band for the night the data were taken. The larger SUPRIME-CAM dataset, where {\it Auriga's Wheel} is found, consists of many overlapping frames taken over the night. From these frames, the overall photometric accuracy is determined to be better than 3.5\% and 2.4\% in the $g$- and $r$-bands, respectively. The airmass correction used was 0.13 mags for the $r$-band and 0.19 mag for the $g$-band (coefficients taken from the Gemini website\footnote{http://www.gemini.edu/sciops/telescopes-and-sites/observing-condition-constraints/extinction}). The Galactic foreground extinction in this region was taken from the interstellar dust maps of \citet{Schlegel1998} and corresponds $E(B-V)$ of 0.1 mag.
\subsection{Spectroscopy}
Upon discovery of the object, 5.6 hours on the Gemini North Telescope were awarded through the Australian Time Allocation Committee to observe the galaxy pair with GMOS-N (Gemini Multi-Object Spectrograph North) in long-slit mode. This allowed for the grating R400 to be used with two position angles (Figure~\ref{slitfig}) and total exposure times of 1 hour with the addition of standard stars for flux calibration. Each one hour exposure was broken into three twenty minute exposures with nodding along the slit to aid in cosmic ray removal. The two slit positions were taken on separate nights, see Table~\ref{obstable} for details. Observations were carried out in queue mode at Gemini-North under Program-ID GN-2009B-Q-87. Calibration-lamp (CuAr) spectra and flat-field frames were provided by GCAL\footnote{Gemini Facility Calibration Unit}. 

The bias subtraction, flat-fielding, wavelength calibration and sky subtraction were executed with the \textsf{gmos} package in the \textsf{gemini} library of the {\sc iraf}\footnote{{\sc iraf} is distributed by the National Optical Astronomy Observatories (NOAO), which is operated by the Association of Universities for Research in Astronomy, Inc. (AURA) under cooperative agreement with the National Science Foundation (NSF).} software. The spectra were extracted from each 2D reduced frame by summing selected CCD lines over our targeted regions. The extracted spectra were obtained by averaging the three individual exposures, using a sigma clipping algorithm to eliminate the effects of cosmic rays. The spectra of the different parts of the ring galaxy have signal-to-noise ratio (SNR) ranging between 5 and 20 and the spectrum of the early type galaxy has a SNR of $\sim$\,60. The full spectral range spans 5200--9400\,\AA, the average spectral resolution is $\Delta\lambda\,=\,4.14$\,\AA\,($\sim\,6.1$ pix.) and the accuracy of the wavelength calibration estimated by measuring the wavelength of 10 lamp emission lines is 0.069\,\AA. 

A spectrum of Feige34 was used for flux calibration and removing the instrument response. Unfortunately, due to the weather conditions, any absolute measurement of the flux is not possible and so the only reliable results from the spectra are related to the equivalent widths of the lines, their flux ratios and the redshift.
\section{Analysis}
\subsection{Photometry}

\begin{figure}
\plotone{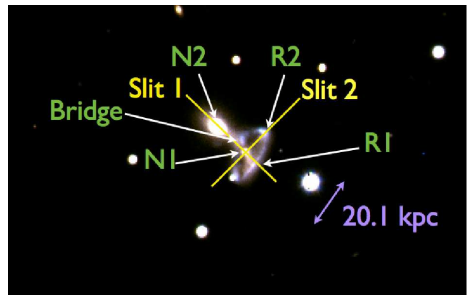}
\caption{ {\it Auriga's Wheel} in pseudo-color showing the position
  angles of the two long-slits used with GMOS-N and the labels given
  to the different regions of the interaction as discussed in this
  paper. N1 refers to the ring galaxy; N2 refers to the early type
  galaxy, R1 is the part of the ring that is on the line between N1
  and N2; R2 is the bright region on the northern side of the ring;
  the Bridge is the region between N1 and N2. A color version of this figure is available in the online version of the article.}\label{slitfig}
\end{figure}

\begin{figure}
\epsscale{.45}
\plotone{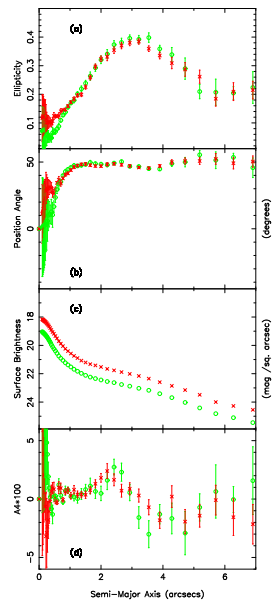}
\caption{Structural parameters of N2 against Semi-Major axis: Ellipticity (a); Position Angle (b); Surface Brightness (c); A4$\times$100 (as per convention) (d). The red crosses are from the $r$-band image and the green circles from the $g$-band image. A color version of this figure is available in the online version of the article.}\label{ellipfig}
\end{figure}

{\it Auriga's Wheel} consists of an early type galaxy (hereafter, N2) seemingly interacting with a disk galaxy (hereafter, N1) dominated by a ring structure. To derive the properties of the early type, we first masked out the disk galaxy and a portion of the bridge region before running the {\sc iraf} routine {\sc ellipse} ~\citep{Jedrzejewski1987} on the early type galaxy. The {\sc ellipse} routine fits elliptical isophotes to the CCD image, allowing the centre, the  position angle and the ellipticity of the elliptical isophote to be free parameters. With this, structural properties of N2 such as half-light radius, surface brightness, as well as evidence for morphological disturbances can be determined. Possible signatures of disturbance could be seen in either a change in position angle of the semi-major axis or more directly with the measure of diskiness/boxiness of N2 from the A4 parameter ~\citep{Bender1988}. Figure~\ref{ellipfig} shows the variation in ellipticity (a), position angle (b), surface brightness (c) and the A4 parameter (d) as a function of the semi-major axis in arcseconds. The red crosses relate the values determined for the $r$-band image and the green circles for the $g$-band image. The ellipticity of the isophotes measured from N2 show it is strongly related to the distance from the centre of the galaxy.  N2 is most elliptical at a galactocentric distance of 3 arcseconds which is where N2 meets the ring. Beyond this distance, N2 becomes more circular. The position angle of the isophotes varies between approximately 45 and 55 degrees at distances larger than 1 arcseconds. Interestingly, this is very close to the angle of the line connecting the centers of the two galaxies, which is about 42 degrees. 

The surface brightness profile of N2 is shown in Figure~\ref{ellipfig}, for the $r$-band (red crosses) and the $g$-band (green points), down to a surface brightness of 25 mag per sq. arcsec. The magnitude of N2 has been determined by considering the total flux inside the last isophote and is found to be 17.29 and 18.22 in $r$- and $g$-bands respectively, after both atmospheric and foreground extinction have been accounted for. This in turn results in a $(g-r)_{o}$ color of 0.93 mag (Table~\ref{galtable}). 

Finally, the A4 parameter ~\citep{Bender1988} can be used as a simple test to relate whether the early type galaxy has undergone an interaction. ~\citet{Khochfar2005} showed that positive A4 values (disky isophotes) are produced by minor mergers involving disk galaxies, while negative A4 values (boxy isophotes) arise from a major merger between two disk galaxies, or between a disk and an early type galaxy. While N2 has mostly positive values of A4 near the core, at galactocentric distances greater than 2 arcseconds the parameter is negative, indicating, together with the variations in ellipticity and position angle, that N2 has been undergoing strong interactions with N1. 

N1,  is a disk galaxy with ring and a nucleus, thus when determining the photometry for this system we have treated the ring and nucleus separately while also considering the galaxy as a whole. To measure the flux coming from the entire disk galaxy we first subtracted the model of the early type galaxy.  This ensured that N2 would not influence the magnitude or color of the disk galaxy determined in this way. A simple ellipse with a semi-major axis of 9.4 arcseconds, semi-minor axis of 4.4 arcseconds and a position angle of $-30$ degrees was placed outside the ring and the flux internal to this was measured. With the zeropoint derived above, we find that the disk galaxy has a total foreground extinction corrected $g$-band magnitude of 18.20 and a $(g-r)_{o}$ color of 0.70 mag. Placing a circular aperture of 1.4 arcseconds solely around the nucleus of N1 we find it has a typical bulge-like color of 1.04 mag and $g$-band magnitude of 20.27 corrected. The color and magnitude of the ring were determined by placing a succession of small square apertures along the ring and fitting to concentric elliptical annuli to it. These methods result in an average color $(g-r)_{o}$  of  0.74.

\subsection{Spectroscopy}

\begin{figure}
\epsscale{.99}
\plotone{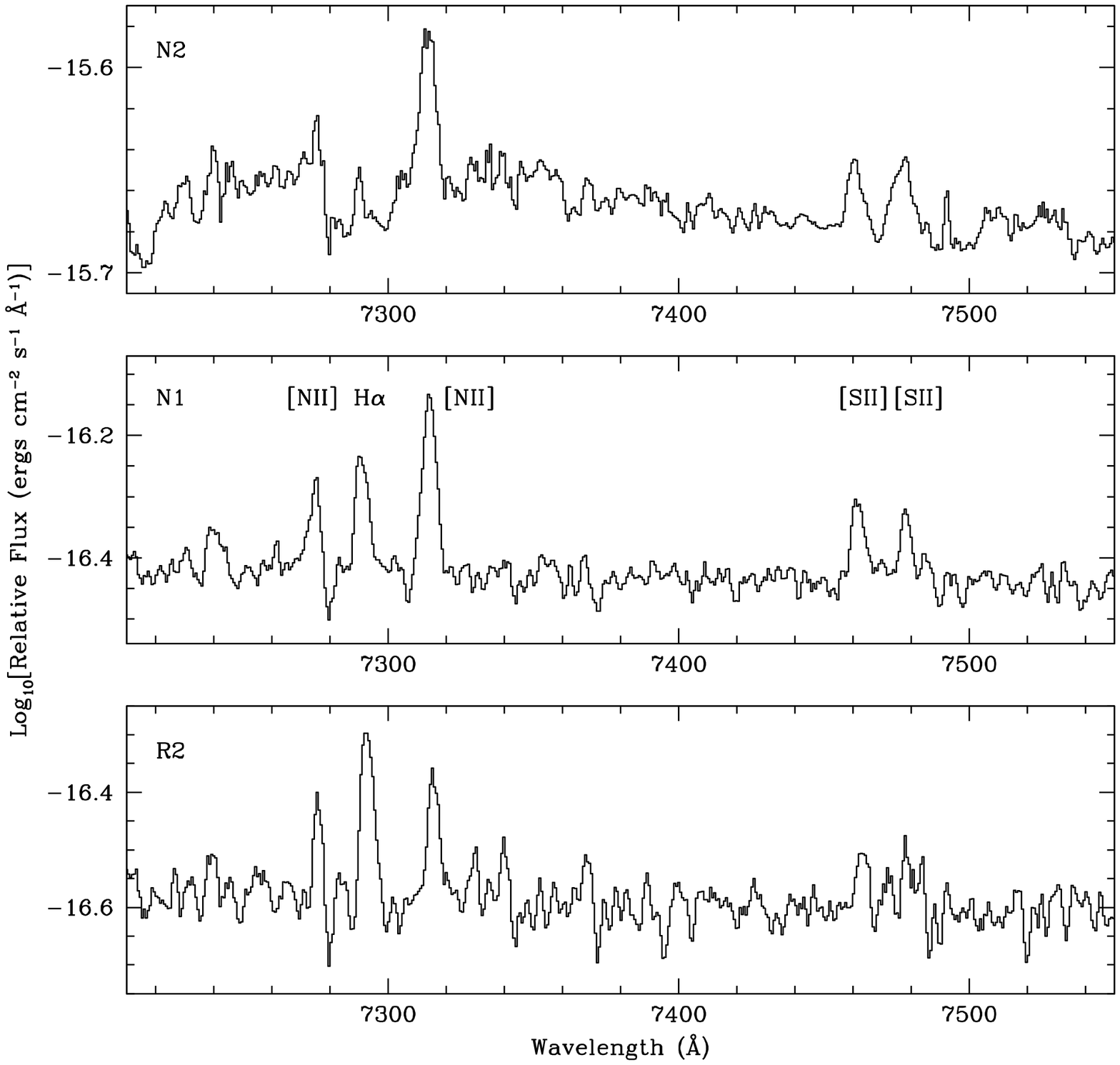}
\caption{Extracted and combined spectra of {\it Auriga's Wheel} in the region of the redshifted emission lines of \ion{N}{2} (6548.1 \AA ~\& 6583.6 \AA), H$\alpha$ (6562.8 \AA) and the \ion{S}{2} doublet (6717 \AA ~\& 6731.3 \AA). The nucleus of the early type galaxy N2 is shown in the top panel, ring galaxy nucleus (N1) in the middle panel and the brightest portion of the ring (R2) in the lower panel. The spectra have been smoothed by a factor of 3.}\label{specfig}
\end{figure}

\begin{deluxetable}{lcc}
\tabletypesize{\scriptsize}
\tablecaption{Properties of the two galaxies.  \label{galtable}}
\tablewidth{0pt}
\tablehead{
\colhead{Properties}& \colhead{N1}&\colhead{N2}
} 
 \startdata
    ¥Total M$_{g}$ (Abs. Mag)& ¥-20.08&-20.06\\
    ¥Total M$_{r}$ (Abs. Mag)& ¥-20.78&-20.99\\	
    ¥Core M$_{g}$ (Abs. Mag)& ¥-18.01&\nodata\\
    ¥Core M$_{r}$ (Abs. Mag)& ¥-19.05&\nodata\\	
    ¥Total $\it{g}$$_{o}$& 18.20&18.22\\ 
    ¥Total $\it(g-r)$$_{o}$& 0.70&0.93\\ 
    ¥Core $\it{g}$$_{o}$ (1.2\arcsec aperture)& 20.27&\nodata\\ 
    ¥Core $\it(g-r)$$_{o}$& 1.04& \nodata\\ 
    ¥Velocity (km s$^{-1}$)& 33131$\pm$33& 33114$\pm$19\\ 
    ¥Redshift (z)& 0.11081& 0.11075\\ 
    ¥$\Delta$z& 0.00011& 0.00006\\ 
    ¥$\Delta$v to N1 (km s$^{-1}$)&\nodata& $-$18\\ 
    ¥$\sigma$ (km s$^{-1}$)\tablenotemark{a}& 235$\pm$16& 290$\pm$16\\ 
    ¥M$\ast$ (M$\odot$)\tablenotemark{b}& 5.18$^{+0.73}_{-0.64}$$\times$10$^{10}$& 11.23$^{+1.58}_{-1.38}$$\times$10$^{10}$\\ 
    ¥M$_{BH}$ (M$\odot$)& 3($\pm$1)$\times$10$^{8}$& 8($\pm$2)$\times$10$^{8}$\\
    ¥$\log\Bigl(\frac{NII}{H\alpha}\Bigr)$& 0.157& $<$ 0.47\\
    ¥Equiv. Width of NII& -5.643& -1.332\\    
    ¥Equiv. Width of H$\alpha$& -3.932& -2.825\\         	
\enddata
    \tablenotetext{a}{Stellar velocity dispersion}
    \tablenotetext{b}{Estimate based on M$_{r}$.}
    \tablecomments{The photometry is accurate to 3.5\% in $g$-band and 2.4\% in $r$-band. The velocity relative to N1 is accurate to $\sim$3 km s$^{-1}$. }

\end{deluxetable}%
The wavelength range of the original spectra with the R400 grating is from 5200 - 9400 Angstroms, however the main region of interest in all the spectra obtained is between the wavelengths of 7200 - 7500 Angstroms.  This is because the redshifted emission lines \ion{N}{2} (6548.1 \AA ~\& 6583.6 \AA), H$\alpha$ (6562.8 \AA) and the \ion{S}{2} doublet (6717 \AA ~\& 6731.3 \AA) as shown in Figure~\ref{specfig} can be found here. While other lines are identifiable in the rest of the spectra, these lines were the clearest and had the best signal to noise. All of the results presented here are based on measurements of these lines. The final redshift determination is the mean redshift from each of the individual emission lines. This interacting pair has a redshift of 0.111. Assuming a Hubble constant of 73 km s$^{-1}$ Mpc$^{-1}$ \citep{Reiss2011}, converting the redshift to distance places these galaxies at 456 Mpc from the Sun after taking into account the Earth's motion around the Sun, the motion of the Sun around the Galaxy and Local Groups' motion towards the Virgo Supercluster. With this distance, the pixel scale of the image is found to be $1'' = 2.2$ kpc.  The semi-major axis of the ring galaxy is then 20.1 kpc and the projected separation between the two nuclei is approximately 13.2 kpc.

\begin{deluxetable}{lccc}
\tabletypesize{\scriptsize}
\tablecaption{Properties of the ring and bridge.\label{ringtable}}
\tablewidth{0pt}
\tablehead{
    \colhead{Properties}& \colhead{R1}&\colhead{R2} & \colhead{Bridge}
    } 
    \startdata	
     ¥Velocity (km s$^{-1}$)& 33302$\pm$30& 33186$\pm$20 & 33063$\pm$26\\ 
    ¥Redshift (z)& 0.11138& 0.11099& 0.11058\\ 
    ¥$\Delta$z& 0.00010& 0.00007& 0.00009\\ 
    ¥$\Delta$v to N1 (km s$^{-1}$)&$+$170& $+$49&$-$68\\ 
\enddata
\tablecomments{$\Delta$v to N1 is accurate to $\sim$3 km s$^{-1}$.}
\end{deluxetable}%

A comparison between the radial velocity at R2 and R1 enables some conclusions on the dynamics of the ring to be drawn. In the following discussion, for simplicity we assume the ring is circular, and inclined to our line-of-sight (l.o.s.). 

If the ring were purely rotating about N1 (see Figure~\ref{slitfig}), then at R2 we would see the large fraction of the ring's rotation velocity directed down our l.o.s. Similarly at R1, the ring's rotation would not be projected down our l.o.s. at all and we would see a zero l.o.s. velocity. In fact, we see a large l.o.s. velocity at R1, ruling out the pure rotation model (Table~\ref{ringtable}). 

If the ring was purely expanding and had no rotation, then we would see a large fraction of the ring expansion velocity projected down our l.o.s. at R1. At R2 the ring would be moving perpendicular to our l.o.s. resulting in a zero l.o.s. velocity. In fact, we observe a non-zero velocity at R2, thus ruling out a pure-expansion model.
 
The simplest explanation is a ring with an element of expansion {\it{and}} rotation. Again, assuming a purely circular ring, whose axis of rotation is inclined at an angle of $i=60^\circ$ to our l.o.s., we can estimate the relative magnitude of the expansion and rotation component. In this scenario, the l.o.s. velocity at R1 is purely due to expansion and has the value $sin(i) v_{\rm{exp}}$. At R2 the l.o.s velocity is purely due to rotation and has the value $sin(i) v_{\rm{rot}}$. This simple analysis would suggest that the ring has a true expansion velocity, $v_{\rm{exp}}\sim$ 197 km s$^{-1}$, and a true rotation velocity, $v_{\rm{rot}}\sim$ 57 km s$^{-1}$. It is therefore dominated by expansion, but with a non-negligible component of rotation.

Additionally, the bridge is seen to be moving towards us, and R1 and R2 are moving away.  We conclude that the bridge represents the near-side of the ring and R1, the far side. The relative velocity separation between N1 and N2 is $-$17.94 km s$^{-1}$, see Table~\ref{galtable}, and suggests N2 is probably in front of N1 and closer to the observer. However, the absolute velocity errors are large and so we can consider the two galaxies being at the same redshift. Better spectra are needed to confidently discriminate the relative motions of these galaxies.

From the emission lines, measuring the logarithm of the ratio of \ion{N}{2} flux over H$\alpha$ flux against the equivalent width of the H$\alpha$ line is an indicator of the type of radiation source exciting these lines.  Stronger \ion{N}{2} emission compared to H$\alpha$ is indicative of an Active Galactic Nucleus  (AGN) while stronger H$\alpha$ emission suggests a starburst origin. Broad equivalent width of the H$\alpha$ line are related to star-forming regions while narrow H$\alpha$ lines are from AGN, see \citet{Cid2010} [their Figure 6]. In our case and in comparison with \citet{Cid2010}, Figure~\ref{specfig} and Table~\ref{galtable} clearly show that in both galactic nuclei, the \ion{N}{2} line is stronger than the H$\alpha$ line and thus we can conclude that a LINER resides in N1 and possibly an AGN in N2.  For the R2 region on the ring, the H$\alpha$ line is stronger than the \ion{N}{2} line and thus, as expected, this region is dominated by star formation \citep{Cid2010}. It seems reasonable that the compression of gas in {\it Auriga's Wheel} has fed the black holes in the centre and thus triggered the AGN phase in both galaxies.

The presence of AGN in a CRG pair is quite rare, only 4 other galaxies in the \citet{Madore2009} sample are considered active and of those all are highly distorted from the interaction with their colliding galaxy.  For those CRGs classified as Seyfert 2 or LINER (IC0614, NGC 0833, NGC 0835, NGC 1142, UGC 5984 ), they all are gas-rich galaxies with significant substructure but none show a ring as clear as {\it Auriga's Wheel}.  It is unclear whether the remaining CRGs have no evidence of AGN or if it has simply not been searched for. While it is entirely possible that the AGN phase in {\it Auriga's Wheel} is unrelated to the collision, N1 is a gas-rich galaxy and so could provide the fuel to power the AGN in both nuclei.

\subsection{Mass estimates}
The color and magnitude information of N1 and N2 can be used to estimate the stellar mass as per Equation~\ref{masseqn} ~\citep{Bell2003}. The luminosity is derived using the zeropoints and redshift converted to Mpc. From this we obtain upper stellar masses for both N1 and N2 of 5.18$^{+0.73}_{-0.64}$$\times$10$^{10}$ and 11.23$^{+1.58}_{-1.38}$$\times$10$^{10}$ M$\odot$ respectively.
\begin{equation}
\log\Bigl(\frac{M\ast}{L\odot}\Bigr) = -0.306 + 1.097(g-r)
\label{masseqn}
\end{equation}

An estimate of the mass of the central black hole in each of the galactic nuclei has been attempted. By cross-correlating against a known galaxy template spectra the stellar velocity dispersion in N1 and N2 can be determined. Table~\ref{galtable} shows the stellar velocity dispersions found for both N1 and N2. In addition, using the approach of \citet{Ferrarese2000}, leads to black hole mass upper estimates of 3$\pm$1$\times$10$^{8}$ M$\odot$ for N1 and 8$\pm$2$\times$10$^{8}$ M$\odot$ for N2 (Table~\ref{galtable}).   

\subsection{Simulation}
\begin{figure}
\begin{center}
\plotone{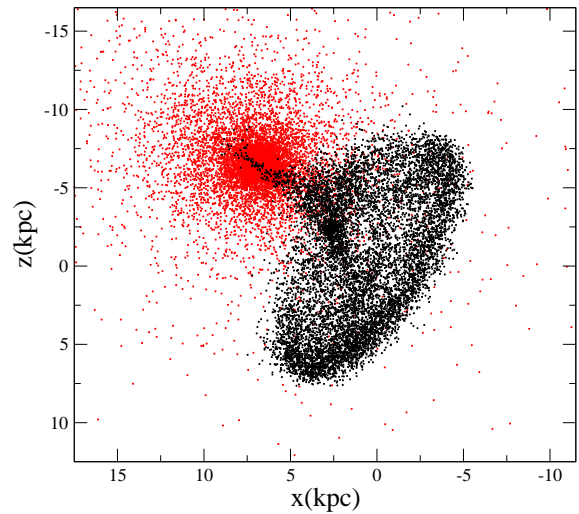}
\caption{x-z plot of star particles from the disk progenitor (black points) and the early type galaxy (red points).}
\label{particlefig}
\end{center}
\end{figure}

Preliminary modeling of the ring galaxy/early type interaction based on Nbody/SPH simulations consists of a spiral galaxy with an exponential disk of stars and gas, embedded in an NFW dark matter halo \citep{NFW1996} and an early type with a Hernquist mass profile \citep{Hernquist1990}. The early type and spiral meet in a near head-on collision. In practice we find that the collision speed must be low ($< 200$ km s$^{-1}$) in order to replicate the observed Bridge-like structure between the two galaxies. Furthermore, we can reasonably reproduce the observed ring morphology if the disk of the spiral is mildly inclined to the collision trajectory. In Figure \ref{particlefig}, we show the final distribution of star particles from the progenitor spiral (black points), and the early type (red points), in our current best-fit model.

We can also account for the observed line-of-sight velocity differences between the nucleus (N1), and the points on the ring (R1, R2, Bridge, see Table~\ref{ringtable}). To accomplish this, the ring is currently expanding rapidly at $\sim$200 km s$^{-1}$ combined with a rotational velocity of $\sim$60 km s$^{-1}$. In the model, this expansion quickly decelerates, confirmation that we are catching the ring shortly after its birth. This phenomena is also seen in the models of Vorobyov \& Bizyaev (2003, their Figure 4). The snapshot in Figure~\ref{particlefig} is $\sim$50 Myr since the collision.

Our current best match to the observations is an encounter between galaxies of approximately equal total mass. We are conducting simulations to investigate the sensitivity of our galaxy-encounter scenario to the mass of the progenitors, and fine-tuning to better match the observed morphology. The full details of the numerical modeling, results and a parameter search will be presented in detail in \citet{Smith2011}.

\subsection{Derived properties}

From the low-redshift catalogue of 127 collision ring galaxies \citep{Madore2009}, after converting the velocity to distance via the Hubble Law and using the separation between the ring and collider in arcseconds, we find that the median projected separation between the ring galaxy and its collider is $\sim$40 kpc. For comparison with {\it Auriga's Wheel}, there are only 11 galaxies for which the calculated projected separation is less than 10 kpc and of these only one (Arp 10) has a pure Cartwheel style morphology. Given the age of the collision in {\it Auriga's Wheel} it would seem that it represents one of the best examples of a clean collisional ring galaxy in formation. Fortunately, despite it's distance, {\it Auriga's Wheel} is bright enough to enable further study.

\section{Conclusion}
 {\it Auriga's Wheel} is an archetypal collisional ring galaxy at a redshift of 0.111. The collision is in its initial stages with the early type colliding galaxy only just having passed the entire way through the disk galaxy. In comparison with the known collisional ring galaxies in the \citet{Madore2009} catalogue,  {\it Auriga's Wheel} shows a very clean interaction with an easily identifiable ring suggesting a near perfect head-on collision scenario. The ring induced from the interaction shows clear evidence of star formation and preliminary modeling shows it to have an expansion velocity of $\sim$200 km s$^{-1}$.  From this we determine that the collision is likely to be around 50 Myr old. Interestingly, both galaxies also contain AGN although it is not clear whether this is due entirely to the collision or simply a property they had beforehand.



\acknowledgments

The authors would like to thank the referee, for helpful comments on improving the paper. BCC would like to thank the European Southern Observatory Fellowship program, the Alexander von Humboldt Foundation Fellowship and the Max-Planck Institute for Astronomy, under which this project was completed. RRL, ANC and RS acknowledge support from the Chilean Center for Astrophysics, FONDAP Nr. 15010003, and from the BASAL Centro de Astrofisica y Tecnologias Afines (CATA) PFB-06/2007. This publication was also financed by the GEMINI-CONICYT Fund, allocated to project No. 32090010. The software package {\sc TOPCAT} (http://www.starlink.ac.uk/topcat/) was used extensively in the preparation of this paper.



{\it Facilities:} \facility{Subaru (SUPRIMECAM)}, \facility{Gemini:North (GMOS-N)}


\end{document}